\documentclass{article}
\usepackage[pdftex]{graphicx}
\usepackage{natbib}
\usepackage[utf8]{inputenc}
\usepackage{authblk}

\usepackage{multirow}
\usepackage{hyperref}
\usepackage{indentfirst}
\usepackage{amsfonts, amssymb, amsmath, amsthm, bm} 
\usepackage{dsfont, accents}

\usepackage{float, enumerate}
\usepackage[mathscr]{euscript}

\usepackage{caption}
\usepackage{subcaption}
\usepackage[section]{placeins}

\title{Reinforcement Learning in Non-Markov Market-Making}
\author[1]{Luca Lalor}
\author[2]{Anatoliy Swishchuk}
\affil[1]{\small University Of Calgary, Department of Mathematics and Statistics, Calgary, AB T2N 1N4, Canada}
\affil[2]{\small University Of Calgary, Department of Mathematics and Statistics, Calgary, AB T2N 1N4, Canada}

\date{\today}

\begin{document}
\maketitle
\thispagestyle{empty}

\begin{abstract}
We develop a deep reinforcement learning (RL) framework for an optimal market-making (MM) trading problem, specifically focusing on price processes with semi-Markov and Hawkes Jump-Diffusion dynamics. We begin by discussing the basics of RL and the deep RL framework used, where we deployed the state-of-the-art Soft Actor-Critic (SAC) algorithm for the deep learning part. The SAC algorithm is an off-policy entropy maximization algorithm more suitable for tackling complex, high-dimensional problems with continuous state and action spaces like in optimal market-making (MM). We introduce the optimal MM problem considered, where we detail all the deterministic and stochastic processes that go into setting up an environment for simulating this strategy. Here we also give an in-depth overview of the jump-diffusion pricing dynamics used, our method for dealing with adverse selection within the limit order book, and we highlight the working parts of our optimization problem. Next, we discuss training and testing results, where we give visuals of how important deterministic and stochastic processes such as the bid/ask, trade executions, inventory, and the reward function evolved. We include a discussion on the limitations of these results, which are important points to note for most diffusion models in this setting.   
\end{abstract}

{\bf Keywords:} Algorithmic and High-Frequency Trading, Limit Order Books, Deep Reinforcement Learning, Hawkes Process, Semi-Markov Process, Market Simulation.

\pagenumbering{arabic}

\newpage

\section{Introduction}

\begin{figure}[H]
	\centering
	\includegraphics[width=0.8\linewidth]{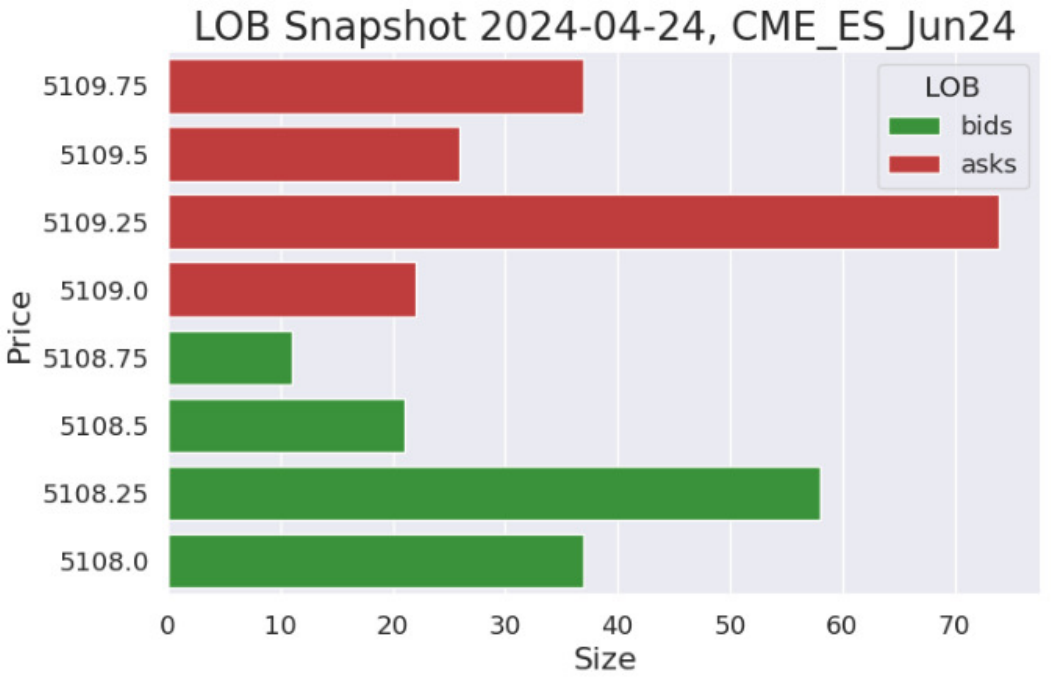}
	\caption{A snapshot of the LOB data on April 24th, 2024 	for the ES futures contract which expired in June 2024, where the x-axis shows the size of the LOs and the y-axis the price.}
	\label{fig:LOB}
\end{figure}

The last 20 years has seen the rising dominance of algorithmic and High-Frequency Trading (HFT) in some of the most liquid financial markets, as this method has become the most common way to complete trade transactions. Transactions in some of the these highly liquid markets, such as equities, futures, or currencies, in particular, occur through the so-called limit order book (LOB) mechanism, which connects the buyers and sellers for the tradeable assets in these financial markets. See figure \ref{fig:LOB}, below, where we show a visual description of the LOB in the E-mini S\&P 500 contract (ES) on April 24th, 2024, which shows a snapshot at a random time during the most active US stock market trading hours (9:30 EST - 16:00 EST). Here one can see the bid and ask prices displayed in the LOB, as well as their respective queue sizes. For a very informed and broad survey on how LOBs are modelled, see \cite{gould2013limit}, and for a more recent survey paper on simulating LOBs see \cite{jain2024limit}, where they highlight many of the major findings from the theoretical and empirical literature on LOBs, as well as addressing many of their limitations, with many of the details in the literature still inadequately formulated. The algorithms used to trade in these markets can place a whole variety of trade order types, where we will specifically focus on strategies that solely implement market and limit orders for simplicity purposes. Extensions can quite often easily be made for the wide array of different trade order types. 

The modeling of stochastic processes in mathematical finance has been a major topic for many years, where now more attention is being placed on modelling the variables in algorithmic and HFT markets. Stochastic Optimal Control (SOC) theory has often been the main method for tackling trading problems in this realm, where \cite{cartea2015algorithmic} formulates a large number of examples deemed relevant for different types of trading needs. Their models often begin with a simple price process following a general arithmetic Brownian motion model. These price processes, however, have been proven to be less well-suited within the algorithmic and HFT setting, as many studies have shown that LOB dynamics often follow non-Markovian properties. One obvious oversight that the literature has found by studying LOB data is that it often experiences jumps i.e., points of discontinuity. The general conclusion from many studies is that LOB models that can portray a dependency in past trade transactions are superior to models with the assumption of an infinitesimal tick size seen in the arithmetic Brownian motion model, as can be seen from the empirical results in \cite{he2019quantitative}, \cite{swishchuk2019compound} and \cite{swishchuk2020general}. Many more studies have provided similar evidence, where some in-depth examples can be found in \cite{cartea2018algorithmic}, \cite{cartea2018enhancing}, \cite{makinen2019forecasting}, \cite{sjogren2021general},  \cite{gavsperov2022deep} and many more. 

As well as improving the price process modeling of LOBs, recent major advancements in Reinforcement Learning (RL), particularly in Deep RL, have enabled the formulation of a more state-of-the-art framework for solving problems in algorithmic and HFT. The most commonly used framework for studying these types of problems has normally been the AS model from \cite{avellaneda2008high}, with numerous examples for various different types of trading problems such as liquidation, acquisition, MM, pairs-trading and statistical arbitrage given in \cite{cartea2015algorithmic} under the SOC setting. These methods, however, are not very robust and an alternative approach that has recently increased in popularity would be to find near-optimal controls under a deep RL framework, which is essentially an approximate solution deviating from the optimal solution by a small amount. 

\begin{figure}[H]
	\centering
	\includegraphics[width=0.8\linewidth]{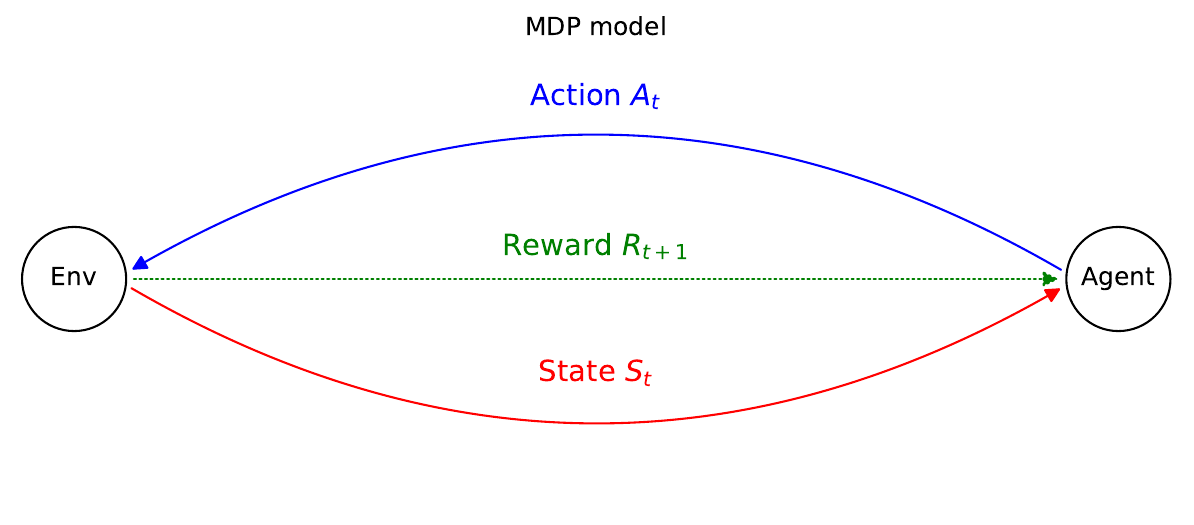}
	\caption{A general MDP model portraying the interaction between the agent and its environment (env).}
	\label{fig:MDP}
\end{figure}

RL, a Machine Learning (ML) algorithm considered to be a mixture of supervised and unsupervised learning, is a technique for solving optimization problems through trial-and-error, where the goal is to maximize some terminal reward. These RL problems are often modelled under a Markov-Decision process (MDP) framework, which is a discrete-time sequential decision-making process where an agent starts with a particular state representation at time $t$, can take a number of actions at time $t$, and based on these actions receives a reward at time $t+1$. See figure \ref{fig:MDP} for a visual description of these interactions between the environment (env) and the agent. 
RL methods, in recent times, have become more compelling as it's now easier to computationally combine them with function approximators such as decision trees and deep neural networks, which has led to significant growth in the innovative field of deep RL. Initial advancements in this area were first made in the groundbreaking success from the work of \cite{mnih2015human}, where they implemented a deep Q-network in the classic Atari 2600 games. From here on out, many more successes were achieved, where some examples can be seen within the game of GO by Google Deepmind in \cite{silver2016mastering}, in text generation in \cite{ranzato2015sequence} and now there are many projects focused on solving the trading problems that occur in todays financial markets. For an in-depth survey on Deep RL we recommend reviewing \cite{arulkumaran2017deep}, for some surveys focusing specifically on their applications to algorithmic and HFT see \cite{pricope2021deep}, and for some specific literature related to optimal MM, the trading problem this paper will focus on, see \cite{gavsperov2021reinforcement}.

The rest of this paper will proceed as follows. We begin in section 2 by briefly introducing some of the basics of the RL and deep RL framework that will be utilized in this paper. We will also highlight here why we believe RL methods trump the standard SOC framework. In section 3 we will outline the MM problem we studied, where we will introduce the processes (deterministic or stochastic) involved and the working parts of the optimization problem. Then, in section 4, we will discuss our training and testing procedures, as well as the results. Here we will portray how certain processes evolved and we will show how our deep RL algorithm was able to learn an optimal trading behavior. We also comment on some of the limitations of these results, as well as how we tried to overcome some of these. This is very important as many models built using much of the standard mathematical finance theory in algorithmic and HFT have often been shown to over-inflate results. Lastly, we give our concluding remarks along with some future research recommendations. 


\section{Deep Reinforcement Learning}

 SOC has often been the standard framework to solve algorithmic and HFT trading problems, with some important works including \cite{bertsimas1998optimal} in dealing with execution costs, \cite{bouchard2011optimal} where a general impulse control approach is applied to algorithmic trading problems, \cite{fodra2015high} applies optimal control to a HFT problem under a Markov Renewal approach, \cite{cartea2015algorithmic} gives a whole range of algorithmic and HFT problems such as acquisition, liquidation, MM, volume imbalance, statistical arbritrage, pairs trading, and in \cite{gueant2017optimal} a thorough theoretical overview is given for the optimal MM problem. In more recent times, the focus has shifted from SOC to deep RL for solving problems in algorithmic trading and HFT. One can find some recent general trading examples with applications in \cite{li2019deep}, \cite{zhang2019deep}, \cite{theate2021application}, and \cite{cartea2021deep}. This shift is mainly due to the inherent advantages in the RL space and we will now discuss three main reasons related to the kind of problems seen in algorithmic and HFT:
 \begin{enumerate}
    \item Model Uncertainty: SOC generally requires a well-defined model of the market dynamics, transition probabilities and reward structure. However, in reality, modeling financial markets is extremely complex due to the presence of many latent variables which could include sentiment, order flow information, liquidity, market microstructure noise, trading interactions and many more. RL does not require a well-defined model of these dynamics, and it can learn optimal strategies from interacting directly with the market. Thus, RL learns to continuously adapt to the markets highly unpredictable stochastic nature, which is essential for developing dynamic systems requiring flexible models. 
     \item Complex and Highly-Dimensional state-action spaces: Traditional SOC models struggle under high-dimensional state representations, which often makes these models intractable. SOC models are also often unable to capture the non-linearity present in financial markets. Deep RL methods, which use function approximators such as decision trees and neural networks, are a lot better at dealing with high-dimensional state-action spaces. RL can also learn complex, non-linear relationships that may be present in the data, which can be very powerful in modelling a market environment that is constantly changing. 
     \item Learning process: In SOC, solutions are normally predefined in the sense that they are computed based on a specific model and then applied without any further adaptation. However, in an environment with constantly changing dynamics, these solutions can often lead to sub-optimal performance. RL algorithms, however, are designed to interact with the environment by exploring the market and exploiting the information obtained by maximizing a cumulative reward function. In this way, the model can discover interesting trading strategies that are not always present in analytical SOC models. Before deploying an RL algorithm in live markets, it can be simulated in a trading environment which will allow it to develop flexible strategies. The RL algorithm can also essentially continuously improve and adapt in real-time to a changing market environment, making it a lot more applicable to live markets. 
 \end{enumerate}

 In this section, we will first briefly introduce the main components of RL in subsection 2.1. Then, in subsection 2.2, we will discuss the framework implemented in our deep RL environment. 

\subsection{Reinforcement Learning (RL)}

Here, we will go through some of the basic components of how one would go about setting up an RL problem. This will be mostly in line with the overviews given in the survey papers by \cite{gavsperov2021reinforcement}, \cite{pricope2021deep} and \cite{singh2022reinforcement}, which focus on introducing RL from a finance/trading/MM perspective. RL problems are formulated through a Markov Decision Process (MDP), as we showed in figure \ref{fig:MDP}, which is a discrete-time sequential decision-making process whereby actions influence not only the immediate rewards, but subsequent decisions i.e., actions influence future rewards and future actions. Here outcomes are partly random and partly influenced by the decision maker. The process involves delayed rewards and the ability to solve the trade-off between immediate and delayed rewards. The main components of the MDP are as follows:
\begin{itemize}
    \item States: A unique characterization of all that is important in a representation of the possible states. This can often be defined as a discrete finite set $S$ in which all states $s \in S$ are possible. In optimal MM, the state space in much of the literature (see table 1 in \cite{gavsperov2021reinforcement}) often includes variables such as inventory, order imbalance, market quality measures, differences between bid/ask prices and many more specific features. 
    \item Actions: Portrays how the system can be controlled. A set of actions $A$, like the state space, can also be defined as a finite set to be applied in each state, denoted as $A(s)$, such that $A(s) \subseteq A$, $\forall s\in S$. A certain number of actions can be taken depending on the state. In optimal MM one can see, again from table 1 in \cite{gavsperov2021reinforcement}, that action spaces in the literature have often included bid/ask price changes, bid/ask size changes, quote pairs, cancelling or posting orders and many more types of actions. 
    \item Transition Function: The system transitions from state to state, based on a probability distribution over the set of possible transitions. The transition function $T$ can be defined as,
    \begin{align}
        T: S \times A \times S \rightarrow [0,1],
    \end{align}
    i.e., the probability of ending up in state $s'$ after completing an action $a$ in state $s$ is denoted as $T(s,a,s')$. Furthermore, $\sum_{s'\in S}T(s,a,s')=1$.
    \item Reward Function: : Rewards are used to determine how the MDP system should be controlled, which can be defined by:
    \begin{align}
        R: S \times A \times S \rightarrow IR,
    \end{align}
    where IR is the immediate reward. The agent should control the system by taking actions that lead to more (positive) rewards over time. In the optimal MM setting, again as shown in table 1 in \cite{gavsperov2021reinforcement}, the reward is normally Profit-and-Loss (PnL) or some utility function like constant absolute risk aversion (CARA) with penalties for variables like inventory. Later on, we will show how we use PnL with a penalty for inventory in our MM setup. 
    \item Policies: Given a quadruple for the MDP, $(S,A,T,R)$,  a policy is a computable function that outputs $\forall s\in S$ an action $a \in A$. Formally, one can define a deterministic policy $\pi$ as a function defined with a mapping $\pi: S \rightarrow A$ or a stochastic policy as $\pi: S \times A \rightarrow [0,1]$ such that $\forall s\in S$, it holds that $\pi(s,a) \ge 0$ and $\sum_{a \in A} \pi(s,a) = 1$. A policy can then be used to evolve an MDP system into making the optimal decisions. 
\end{itemize}

Then to maximize the accumulated reward over time, the agent learns to select her actions based on her past experiences (exploitation) and/or trying new choices (exploration). There is a trade-off between exploration and exploitation and it’s crucial in designing the RL algorithm to improve learning and performance. There are many works, such as \cite{berger2014exploration}, that develops a multidisciplinary framework for dealing with this issue. 

Next, to solve a RL problem, we need to find a policy that obtains the largest reward over a certain period. Constructing an optimal model is a necessary step in this regard. This often has two steps: the goal of the agent, what is being optimized, and the second step is the optimal way in which the goal is being optimized. To this end, first define the discounted sum of rewards as:
\begin{align}
    R_t^{\gamma} = \sum_{k=0}^{T-t} \gamma^k R_{t+k} = R_t + \gamma R_{t+1} + \gamma^2 R_{t+2}+...+\gamma^{T-t}R_T.
\end{align}
Here, $T < \infty$ and $\gamma \in [0,1]$ is a discount factor which plays an important role in future rewards. The goal of the discounted average reward in an MDP is to find a policy $\pi^*$ that maximizes the expected return $\mathbb{E}_{\pi}[R_t^{\gamma}]$. Then, the state value function $V^{\pi}(s)$ of an MDP is the expected reward starting from state $s$, and then following policy $\pi$. This can be represented as,
\begin{align}
    V^{\pi}(s) = \mathbb{E}_{\pi}[R_t^{\gamma}|S_t=s] = \mathbb{E}_{\pi}\left[ \sum_{k=0}^{T-t} \gamma^k R_{t+k}|S_t=s\right].
\end{align}
Next, the state-action value function, which we will define as $Q^{\pi}(s,a)$, is the expected reward starting from state $s$, taking action $a$, and then following policy $\pi$. This can be represented as,
\begin{align}
    Q^{\pi}(s,a) = \mathbb{E}_{\pi}[R_t^{\gamma}|S_t=s, A_t=a] = \mathbb{E}_{\pi}\left[ \sum_{k=0}^{T-t} \gamma^k R_{t+k}|S_t=s, A_t=a\right].
\end{align}
While $\pi$ can be a policy, $\pi^*$ denotes the optimal one with the highest expected cumulative reward i.e., $V^{\pi^*} (s) \ge V^{\pi} (s), \forall s\in S$ and $\forall \pi$. So,
\begin{align}
    V^{\pi^*}(s) = \sup_{\pi} V^{\pi}(s),
\end{align}
and, similarly, the optimal Q-value is,
\begin{align}
    Q^{\pi^*}(s, a) = \sup_{\pi} Q^{\pi}(s, a).
\end{align}

To briefly summarize, in an RL system, input and output pairs are not provided. Instead, the system is a given a specific goal, a set of allowable actions and environmental constraints for their outcomes. The agent interacts with the environment through trial and error and learns to optimize the maximum reward. Popular RL algorithms use equations (3) and (4) above to estimate the sum of the discounted rewards, where the function is defined by a tabular mapping.

\subsection{Deep Reinforcement Learning}

In this subsection, we will give a brief overview on how to implement deep learning techniques in an RL environment and how it can lead to improved and more robust solutions to some of the more complicated optimization problems, like an optimal MM problem. One of the more major recent advancements that has increased the popularity of RL models nowadays is the use of deep RL, where a function approximator, such as a neural network or a decision tree, can estimate the states as explained in many works and more recently in \cite{singh2022reinforcement}. Deep RL uses function approximation instead of tabular methods to estimate the state values. Functional approximation eliminates the need to store all state and value pairs in a table and enables the agent to generalize the value of states it has never seen before or states the agent has partial information about by using the value of similar states. So, to reiterate, RL dynamically learns with trial-and-error methods to maximize the rewards in a tabular format, while deep RL combines RL with neural networks to tackle high-dimensional state and action spaces where it is too difficult to solve the problem using the standard RL tabular format.

There are three specific deep RL approaches that are often used in the academic literature within an optimal MM problem and we will focus on them here as we use them from section 3 onward. These three approaches include the Critic-only, Actor-only, and Actor-Critic approaches. In our optimal MM problem, we deployed the Actor-Critic method. This method has seen a few applications in the optimal MM literature, with most being very recent. Examples include \cite{chan2001electronic},  \cite{gueant2019deep}, \cite{sadighian2019deep}, \cite{sadighian2020extending}, \cite{gavsperov2022deep}, and \cite{baldacci2023market}. In order to understand this method, we first give a brief overview of the Critic-only and Actor-only approaches, as these are the foundations of the Actor-Critic approach. And so, these three methods can be summarized as follows,
\begin{itemize}
    \item Critic-only: This is the most common approach in the literature. Under this model, the goal is to learn the value function where the agent can learn the expected outcomes of the different actions. Then, during the decision-making process, the agent senses the current state of the environment and selects the action with the best outcome according to the value function. The reward function in the critic-only approach does not need to be differentiable and is highly flexible, making it applicable to a comprehensive set of problems. Additionally, this property allows modeling complex reward schemes. Furthermore, the preference between immediate and future rewards can be carefully controlled due to the explicit use of a discount factor (see equation (3)). However, the most noticeable limitation is the agents discrete action space closely related to Bellmans “curse of dimensionality”. To overcome this, an effort to enrich the state with other data sources and whether immediate or terminal rewards perform better and under what experiments can be explored.
    \item Actor-only: In this approach, the agent views the state of the environment and acts directly, i.e., without computing and comparing the expected outcomes of different actions. Hence, the agent learns a direct mapping (a policy) from states to actions. The main advantages are its continuous action space, faster convergence, and higher transparency. Having continuous actions, the agent can carefully interact with the environment, for example, to gradually increase an investment. The noticeable disadvantage with this approach is that the actor-only approach needs a differentiable reward function, limiting the reward schemes that can be modelled. However, the impact of different neural network architectures for deep RL agents and the effects of varying reward functions can help overcome this.
    \item Actor-Critic: This approach combines the actor-only and critic-only RL approaches and contains two agents, the actor and the critic. The actor determines the actions and shapes the policy of the system. At each step, the actor takes the current state as an input and computes the agent’s action as output. The critic assesses these actions. Thus, it gets the current state and the actor’s action as input and computes the discounted future rewards as output. The principal goal is to steadily modify the actor’s policy parameters to maximize the reward predicted by the critic. Despite the ambition to combine the advantages of both agents, only a few studies are employing actor-critic RL in financial markets.
\end{itemize} 

In this paper we used the Soft Actor-Critic (SAC) algorithm, which was first developed by \cite{haarnoja2018soft}, and was also used in the deep RL optimal MM problem studied in \cite{gavsperov2022deep}. The SAC algorithm is an off-policy (a continuously adaptable policy) maximum entropy based deep RL algorithm, where the actor tries to maximize entropy and the expected reward. Examples of similar RL algorithms constructed under the maximum entropy framework can be found in \cite{ziebart2008maximum}, \cite{toussaint2009robot}, \cite{rawlik2013stochastic}, \cite{fox2015taming} and \cite{haarnoja2017reinforcement}. Incorporating entropy into the reward function essentially encourages exploration. The key components of the algorithm, as explained in \cite{haarnoja2018soft}, can be outlined as follows:
\begin{itemize}
    \item Policy update: The policy, referred to as the actor, is updated by minimizing the following objective function:
    \begin{align}
        J_{V}(\psi) = \mathbb{E}_{S_t\sim \mathscr{D}}\left[\frac{1}{2}\left(V_{\psi} (S_t)-\mathbb{E}_{a_t\sim \pi_{\phi}}\left[Q_\theta (S_t, a_t)-log \pi_{\phi}(a_t|s_t)\right]\right)^2\right],
    \end{align}
    where $\psi$, $\theta$ and $\phi$ are the parameters of the respective neural networks, $V_\psi (S_t)$ is the state value function, $Q_\theta (s_t, a_t)$ is a soft Q-function, $\pi_{\phi}$ is a tractable policy, and $\mathscr{D}$ is the replay buffer, also referred to as the distribution of the earlier sampled states and actions. An unbiased estimator can then be used to estimate the gradient as follows,
    \begin{align}
        \hat{\nabla}_\psi J_V(\psi) = \nabla_\psi V_\psi (S_t)(V_\psi (S_t)-Q_\theta (s_t, a_t) + log \pi_\phi (a_t|s_t)),
    \end{align}
    where the sampled actions come from the current policy. 
    \item Q-Value update: The Q-value function, referred to as the critic, is updated by minimizing the following soft Bellman residual,
    \begin{align}
        J_Q(\theta) = \mathbb{E}_{(s_t, a_t)\sim \mathscr{D}}\left[\frac{1}{2}(Q_\theta (s_t, a_t)-\hat{Q}(s_t, a_t))^2\right],
    \end{align}
    where the target value $\hat{Q}(s_t, a_t)$ is given by,
    \begin{align}
        \hat{Q}(s_t, a_t) = r(s_t, a_t) + \gamma \mathbb{E}_{s_{t+1} \sim {\pi_\theta}}[V_{\hat{\psi}}(s_{t+1})].
    \end{align}
    This, again, can be computed using stochastic gradient descent as follows, 
    \begin{align}
        \hat{\nabla}_\theta J_Q (\theta) = \nabla Q_\theta (a_t, s_t)\left(Q_\theta (a_t, s_t) - r(a_t, s_t)- \gamma V_{\hat{\psi}}(s_{t+1})\right).
    \end{align}
    To ensure stability in the training process, the update utilizes the target network $V_{\hat{\psi}}$, where $\hat{\psi}$ is an exponentially weighted moving average of the network. 
    \item Policy parameters: The parameters can be adjusted to control the entropy of the policy by directly minimizing the expected Kullback-Leibler (KL)-divergence, also known as relative entropy. Using a neural network transformation, it is then convenient to re-parameterize the policy by setting $a_t = f_\phi (\epsilon_t; s_t)$, where $\epsilon_t$ is a noise term which can be sampled from a Gaussian distribution. Then the policy parameters can be learned as follows, 
    \begin{align}
        J_\pi (\phi) = \mathbb{E}_{s_t \sim \mathscr{D}, \epsilon_t \sim \mathscr{N}}[log \pi_\phi (f_\phi (\epsilon_t; s_t)|s_t)-Q_\theta(s_t, f_\phi(\epsilon_t;s_t))].
    \end{align}
    Here, $\pi_\phi$ is defined implicitly in terms of $f_\phi$. The gradient can next be approximated as,
    \begin{align}
    \begin{split}
        (\hat\nabla_\phi)J_\pi (\phi) =& \nabla_\phi log \pi_\theta (a_t, s_t) + (\nabla_{a_t} log \pi_\phi (a_t|s_t) \\&-\nabla_{a_t}Q(s_t, a_t))\nabla_\phi f_\phi(\epsilon_t;s_t).
    \end{split}
    \end{align}
    Here, $a_t$ is evaluated at $f_\phi(\epsilon_t; s_t)$ as previously stated.     
\end{itemize}

Thus, to conclude, the SAC method combines the actor and critic method to leverage the benefits of both, specifically focusing on improving exploration through entropy maximization. The balance between exploration and exploitation is controlled by all the parameters. 



\section{Optimal Market-Making Problem}

A MM problem, in an optimization sense, involves a financial market player who would like to maximize their terminal wealth by frequently trading in and out of positions, often using limit orders. These types of traders are also regularly known as liquidity providers, as placing a large amount of limit orders in the LOB is seen as providing liquidity to the market. Similarly to the standard SOC MM setting, there is a variable the agent can control, which is normally referred to as an action in the RL setting. In our optimal MM problem, the control is whether or not to be posted at the best bid/ask in the LOB. To summarize this section, in section 3.1 we will first describe the key stochastic processes that go into creating a model that can be used to simulate a LOB in a MM setting. Then, in section 3.2, we describe the deep RL framework we developed to perform our analysis i.e., our method for solving the optimal MM problem.

\subsection{Optimal Market-Making Model Dynamics}

Here we will first describe the key stochastic processes that form our optimal MM problem, formulated similarly to the processes given in \cite{cartea2015algorithmic} and in \cite{lalor2024market} for a similar MM problem, where there they are utilized under the standard MM framework. These processes are as follows:
\begin{itemize}
    \item $A=(A_t^{\pm})_{\{0 \le t \le T\}}$, where $A_t^{\pm} \in \{-1,0, 1\}$, refers to the actions the agent can take (sell, hold and buy), also known as the agents control process. This tells the market-maker whether or not to be posted at the best bid/ask.
    \item $Q^{A}=(Q_t^{A})_{\{0 \le t \le T\}}$ is the agents' controlled inventory process and is impacted by how much the agent trades. 
    \item $P=(P_t)_{\{0 \le t \le T\}}$ is the midprice process of the financial asset being traded. In our setup we assume the MM trades small enough so that their influence on the midprice is negligible. 
    \item $M^{\pm}=(M_t^{\pm})_{\{0 \le t \le T\}}$ is a counting process for the arrival of market orders into the market. Here, the $\pm$ indicates the arrival of buy and sell market orders, respectively. 
    \item $N^{A, \pm}=(N_t^{A, \pm})_{\{0 \le t \le T\}}$ is the trade order fill process, which is also a counting process but with a dependence on the LO posting control. Here, the $\pm$ indicates limit orders that are filled as they are matched to incoming market orders, $M^{\pm}$. 
    \item $C^{A}=(C_t^{A})_{\{0 \le t \le T\}}$ is the agents cash process, which is essentially a running profit and loss function, as a result of executing the strategy. 
\end{itemize}

Next, we show how these processes satisfy certain differential equations, which may be stochastic, as follows:

\begin{itemize}

    \item For the price process, we follow the method in \cite{lalor2024algorithmic}, where their price processes follow semi-Markov and Hawkes jump-diffusion dynamics. This type of price process follows the non-Markovian dynamics normally seen in LOB data, as shown in many studies such as in  \cite{cartea2018enhancing}, \cite{cartea2018algorithmic}, \cite{he2019quantitative}, \cite{swishchuk2019compound}, \cite{swishchuk2020general}, and many more. Thus, we believe our method here to be a more accurate way to formulate the price process. A general version of this type of price process for an optimal MM problem can be described as follows,
    \begin{align}
    dP_t = \eta dt + \sqrt{\sigma^2+\bar{\sigma}^2 + \varsigma^2} dW_t, 
    \end{align}
    Here, $\eta$, $\bar{\sigma}$ and $\varsigma$ can either be represented for the semi-Markov case (which we could then denote as $\eta_{SM}$, $\bar{\sigma}_{SM}$ and $\varsigma_{SM}$) or for the Hawkes case (which we could then denote as $\eta_{HP}$, $\bar{\sigma}_{HP}$ and $\varsigma_{HP}$). This price process was formulated via a diffusion approximation, as shown in \cite{lalor2024algorithmic}, where the original theoretical frameworks for these approximations are given in \cite{swishchuk2017semi} for the semi-Markov case and in \cite{swishchuk2020general} for the Hawkes case. We will now briefly summarize how each of the parameters in equation(15), for the semi-Markov and Hawkes cases, can be described:
    \begin{itemize}
        \item $\eta_{SM}$: Here we define this for the balanced market case, as first shown in \cite{roldan2023stochastic},  as follows,
        \begin{align}
        \eta_{SM} = \frac{1}{m_{\tau}}s^*, 
        \end{align}
        where $s^*$ represents a formulation for a normalized process and is defined as $s^*:=\delta(2\pi^*-1)$, where $\pi^*$ is a long-run probability. Here, $\delta$ represents the tick size and $m_{\tau}$ can be defined for a two state Markov chain as,
        \begin{align}
            m_\tau := \sum_{i\in \{-\delta,\delta\}}\pi^*(i)m(i).
        \end{align}
        \item $\bar{\sigma}_{SM}$: Can be as defined in \cite{swishchuk2017semi} as follows:
        \begin{align}
        \bar{\sigma}_{SM} = \sqrt{\frac{(\sigma^*)^2}{m_{\tau}}+\frac{\Pi\sigma^2}{m_{\tau}}}.
        \end{align}
        Here, $\sigma^*$ is defined as,
        \begin{align}
        \sigma^* = \sqrt{4\delta^2\left(\frac{1-p_{cont}^{\text{'}}+\pi^*(p_{cont}^{\text{'}}-p_{cont})}{(p_{cont}+p_{cont}^{\text{'}}-2)^2}\right)},
        \end{align}
        where $p_{cont} = P[X_{k+1} = \delta | X_k = \delta]$, $p_{cont}^{\text{'}} = P[X_{k+1} = -\delta | X_k = -\delta]$, $\pi^* = \frac{p_{cont}^{\text{'}}-1}{p_{cont}+p_{cont}^{\text{'}}-2}$, $\tau = \sum_{k=1}^{\infty} \sum_{p=1}^{\infty} \alpha^b(k)\alpha^a(p)f^*(k,p)$ and $f^*(k,p) = \pi^*f(k,p)+(1-\pi^*)\tilde{f}(k,p)$, where $f(k,p)$ is the probability distribution after a price increase and $\tilde{f}(k,p)$ is the probability distribution after a price decrease and $\alpha$ refers to the function of the inter-arrival times, where the exponents $a$ and $b$ refer to the ask and bid sides of the LOB, respectively (see  \cite{swishchuk2017semi} for more details).
        \item $\varsigma_{SM}$: The coefficient representing the the Semi-Markov diffusion approximation for the jump part, can be defined as in \cite{lalor2024algorithmic},
        \begin{align}
            \varsigma_{SM} = \frac{\sigma^*}{\sqrt{\tau}}.
        \end{align}
        Here $\sigma^*$, as defined in equation (19) is a constant depending on the ergodic and transition probabilities of a Markov chain and $\tau$ refers to the inter-arrival time of the jumps.
        \item $\eta_{HP}$: For the Hawkes case, the drift coefficient in equation (15) can be defined, as in \cite{roldan2022optimal} and \cite{roldan2023stochastic}, as,
        \begin{align}
        \eta_{HP} = a^*\frac{\lambda}{1-\hat{\mu}},
        \end{align}
        where $a^*$ is a constant depending on the transition probabilities of a Markov chain. This can be defined as in \cite{swishchuk2020general} as $a^*:= \sum_{i \in X}\pi_i^*a(i)$, where $\pi^*$ are the ergodic probabilities of a Markov chain, which here is defined as $X$. $\lambda$ is the background intensity and $\hat{\mu}$ refers to the output of the response/excitation function of the Hawkes process.
        \item $\bar{\sigma}_{HP}$: can be defined as in \cite{swishchuk2020general} as,
        \begin{align}
        \bar{\sigma}_{HP} = \sqrt{(\sigma^*)^2+\left(a^*\sqrt{\frac{\lambda}{1-\hat{\mu}}}\right)},
        \end{align}
        where,
        \begin{align*}
        \sigma^* = \hat{\sigma}\sqrt{\frac{\lambda}{1-\hat{\mu}}}\ \text{and} \ \hat{\sigma}^2 := \sum_{i \in X}\pi_i^*v(i), 
        \end{align*}
        and $v(i)$ represent the transitions in and out of states.
        \item $\varsigma_{SM}$:  Can be as defined in \cite{lalor2024algorithmic} as, 
        \begin{align}
            \varsigma_{HP} = \sigma^*\sqrt{\frac{\lambda}{1-\hat{\mu}}}. 
        \end{align}
        Here, $\sigma^*$ is again a constant depending on the ergodic probabilities of the Markov chain.
    \end{itemize}
    Note, we remove the indices SM and HP, as in equation (15), as the mathematical formulations are the same for both cases from here on out. Although still situational, we generally believe that the Hawkes process dynamics are more suitable for modelling LOB data, based on much of the empirical evidence in the LOB data portraying event clustering and self-excitation dynamics. This was shown in many of the works we highlighted in the introduction. Three of the main differences in the semi-Markov and Hawkes process models relates to how it it models memory, transition dynamics and its intensity function. Firstly, due to the limited time spent in a state, semi-Markov processes have a restricted memory, while the memory in Hawkes processes includes all the past events. Secondly, the state transitions in a semi-Markov processes have arbitrary holding times, while the event arrival times in a Hawkes processes are based on previous events. Thirdly, transitions influence the distribution of the holding times in semi-Markov processes, while the intensity function for Hawkes processes controls the arrival rate, which is conditional on past events. 
         
    \item The controlled inventory process keeps track of the agents position in the market. This alters whenever the agent has a limit order posted in the market and proceeds to get filled by an incoming market order. This process thus satisfies,
    \begin{align} 
    Q_t^{A}=N_t^{A,-}-N_t^{A, +},
    \end{align}
    where $N_t^{A,-}$ ($N_t^{A,+}$) indicate buy (sell) limit order fills that increases (decreases) the agents inventory. In order to calculate whether buy or sell limit order fills occurred, we follow the method in \cite{lalor2024market}. In their studies, they split up trade order fills into adverse and non-adverse fills, as well as placing a probability on non-adverse fills occurring. More specifically, non-adverse fills are defined discretely as, 
    \begin{align}
    NFA_t = \sum_{i=1}^NA^+_{t_i}\mathbb{I}_{\{M^+_{t_i}=1\}}*p,
    \end{align}
    and
    \begin{align}
    NFB_t = \sum_{i=1}^NA^-_{t_i}\mathbb{I}_{\{M^-_{t_i}=1\}}*p,
    \end{align}
    for $i = 0,...,N$, where $N$ is the total number of time steps. Here, $NFA_t$ and $NFB_t$ are the counting processes for all non-adverse fills on the best ask and best bid, respectively. $p$ here represents what's called a non-adverse fill probability, which can be defined as:
    \begin{align}
    p = P(N_{t_i}^{A, \pm}|A_{t_i}^+\mathbb{I}_{\{AS(t_i)\ge AS(t_{i+1})\}}=0, A_{t_i}^-\mathbb{I}_{\{BS(t_i)\le BS(t_{i+1})\}}=0).
    \end{align}
    This non-adverse fill probability represents the probability of getting filled given the fill is non adverse, where $BS_{t_i}$ and $AS_{t_i}$ indicate the bid and ask prices at time $t_i$, respectively. In \cite{lalor2024market} they also keep track of the adverse fills discretely in the following manner:
    \begin{align}
    AFA_{t} = \sum_{i=1}^NA_{t_i}^+\mathbb{I}_{\{AS(t_i)<AS(t_{i+1})\}},
    \end{align}
    and
    \begin{align}
    AFB_{t} = \sum_{i=1}^NA_{t_i}^-\mathbb{I}_{\{BS(t_i)>BS(t_{i+1})\}},
    \end{align}
    for $i = 0,...,N$. Here $AFA_t$ and $AFB_t$ represent the counting processes for all adverse fills that occur on the best ask and bid, respectively. Intuitively, this can be interpreted as follows: every time the MM strategy has a limit order posted on the best bid (ask) $A_{t_i}^-(A_{t_i}^+)$ and the asset price at time step $t_{i+1}$ is lower (higher) than at time $t_i$, the traders limit order was then filled at the price $BS_{t_i} (AS_{t_i})$. In real live markets operating under the LOB system, the price of any asset can not move below (above) a traders posted bid (ask) limit order without filling their limit order first, thus this makes sense based on the rules of the LOB system. As well as the analysis given in \cite{lalor2024market}, another thorough overview of the adverse fill problem can also be found in \cite{delise2024negative}, where they refer to this phenomena as the "negative drift of a limit order fill". The rest of the trade order fills the MM receives are considered non-adverse fills. Lastly, it is necessary to combine the two fill tracking formulas on both sides of the LOB in order to determine whether a trade order fill occurred or not, which can be done as follows,
    \begin{align}
    N_{t}^{A, +} =  \sum_{i=1}^N max(AFA_{t_i},NFA_{t_i}),
    \end{align}
    and 
    \begin{align}
    N_{t}^{A, -} =  \sum_{i=1}^N max(AFB_{t_i},NFB_{t_i}).
    \end{align}
    Here $N_{t}^{A, +}$ and $N_{t}^{A, -}$ are now counting processes collecting all the trade order fills that occur throughout this particular trading strategy.
    \item The agent's cash process satisfies the SDE,
    \begin{align}
    dC_t^{A}=\displaystyle{\left(P_t+\frac{\Delta}{2}\right)}dN_t^{A, +}-\displaystyle{\left(P_t-\frac{\Delta}{2}\right)}dN_t^{A, -}.
    \end{align}
    Here, $\Delta$ is the spread which is the difference between the best bid and ask, and, as before, $N_t^{A, \pm}$ represents the counting process for filled LOs.
\end{itemize}

\subsection{Deep Reinforcement Learning Framework}

In this subsection we will describe the important working parts of the optimization problem we created to solve our optimal MM problem. Parts of this are similar to the deep RL setup given in \cite{gavsperov2022deep}.

We first begin by defining our state space as follows:
\begin{align}
    S_t = (P_t, Q_t^A).
\end{align}
Recall, $P_t$ is the midprice process and $Q_t^A$ is the controlled inventory process. For the inventory process, we set a constraint whereby the agent has a maximum position size $q$, thus there are $2q+1$ possible inventory states such that $Q_t^A \in \{-q,...,q\}$. In order to make the inputs to our Neural Network more balanced later on, we normalize the midprice process $P_t$ and inventory process $Q_t^A$ using z-score and min-max normalization, respectively.   

Next we define the action space for a setting where the agent can post buy, hold and sell trade orders, which is the part of the system the agent controls. This is defined as follows,

\begin{align}
A_t = 
\begin{cases} 
\{0, 1\}, & \text{if } Q_t^A = -q, \\
\{-1, 0, 1\}, & \text{if } -q < Q_t^A < q, \\
\{-1, 0\}, & \text{if } Q_t^A = q. \\
\end{cases}
\end{align}
Recall that in this particular MM problem, the agents possible decisions (i.e. actions) is whether or not to post a limit order at the best bid/ask in the LOB and this must be within the inventory constraints defined earlier. The quantity of each limit order that ends up being posted in our simulations is assumed to be of size $1$. 

In order to measure the performance of the optimal MM strategy, we must setup a function that calculates the rewards. In our optimal MM problem we follow the same reward function as in \cite{gavsperov2022deep}, which is also very similar to the value functions used under the standard MM trading problem frameworks in \cite{cartea2015algorithmic}. The reward assumes that the MM aims to maximize the expression,
\begin{align}
    \mathbb{E}_\pi \left[W_T^A-\alpha\int_0^T|Q_t^A|dt\right],
\end{align}
which aims to select the best available policy in the set of possible policies $\pi$. As defined in section 2.1, $\pi$ is a mapping from states to actions and in our case this is of course a stochastic policy. Here, the total wealth is given by $W_t^A = Q_t^A P_t + C_t^A$, and $\alpha\ge 0$ is a running inventory penalty that penalizes nonzero inventories. 

Next, we will describe the implementation of the neural networks within this deep RL optimal MM problem. As in \cite{gavsperov2022deep}, we deploy the SAC method as described in section 2.2 and we also use multi-layer perceptron networks. Recall that here there is a network for the actor, the critic and for extracting the parameters as described in equations (8), (10) and (13). We will now summarize how each of these neural networks are formulated:
\begin{itemize}
    \item Actor Neural Network: Given our state space defined in equation (33), this neural  network is designed to the predict the mean and log standard deviation of a Gaussian distribution for the set of possible actions given in equation (34). Thus, our input layer into the neural network is the state vector $\bf{s} \in \mathbb{R}^n$, where $n=2$ for our 2-dimensional state space. Then our neural network has two hidden layers, each with 256 neurons. The first hidden layer can be defined as $h_1=ReLU(W_1s+b_1)$ and the second hidden layer as $h_2=ReLU(W_2h_1+b_2)$, where $W_1$ and $W_2$ are the neural network's weights, and $b_1$ and $b_2$ are the bias terms. The activation function used, ReLU, is called rectified linear units and is a popular choice for avoiding vanishing gradients. The output layer consists of one part for the mean and one for the log standard deviations of the action distribution. The mean output is defined as $\mu = W_\mu h_2+b_\mu$ and the log standard deviation output layer is defined as $log \sigma = W_{log (\sigma)}h_2 + b_{log(\sigma)}$.
    \item Critic Neural Network: To estimate the state-action value functions, the SAC method uses two critic neural networks, $Q_1$ and $Q_2$, in order to reduce the overestimation bias often seen in single critic methods, where each critic neural network has an almost identical architecture to the ones just described for the actor neural network. Its input is the state-action pair $[s,a] \in \mathbb{R}^{n+m}$, where here $n=2$ and $m=1$ for our state and action dimensions, respectively. The hidden layers, as for the actor neural network, are $h_1 = ReLU(W_1[s,a]+b_1)$ and $h_2 = ReLU(W_2h_1+b_2)$. The output layer, representing the state-action Q-values, is defined as $Q(s,a) = W_Qh_2+b_Q$. 
    \item Parameter Neural Network: To extract the parameter values, an almost identical neural network is again used here. The input to the neural network is the state vector $\bf{s}$, the hidden layers are $h_1 = ReLU(W_{h_1}s+b_{h_1})$ and $h_2 = ReLU(W_{h_2}h_1+b_{h_2})$, and the output of this network is used as the input to both the actor and critic networks defined above. 
\end{itemize}
Our training is then conducted using this SAC method, where we will show some results in the next section. 

\section{Results}

In order to perform our analysis we first trained our model on our simulated data, which was constructed based on the analysis in section 3. Please see below, in table \ref{tab:table1}, the parameter values used. The majority of these parameters were purely picked for illustrative purposes to assess how the deep RL algorithm could learn and perform, although some strategic selections were made. We kept the time step small as we would like to assess the strategy from a HFT standpoint. We set the maximum inventory parameter $q=5$ to avoid large positions significantly influencing the results. We will show results where we picked $20\%$ for our non-adverse fill probability, which is in line with empirical results in \cite{lalor2024market} for estimating this probability based on some of the most liquid futures contracts listed on the Chicago Mercantile Exchange. We will then also proceed to show two sets of results, one with adverse fills and one without, in order to portray how significant the effect adverse fills can be on the overall results. We set the spread between the bid and the ask to $\Delta=0.01$ which is quite common throughout the majority of the day in many liquid financial products and we set the running inventory penalty to $0.001$, which had quite a bit of emphasis on the test results as inventory is generally very low. For our in-sample results, we set the number of simulations to $10^6$ for the training set, which lead to us training 1000 episodes. For our out-of-sample results, we then tested our training results on 200 unseen episodes. Computationally, we performed our analysis on python and for the SAC analysis part we made use of the Stable Baselines3 python package, developed by \cite{stable-baselines3}, which was also used by \cite{gavsperov2022deep} in their optimal MM problem. 

\begin{table}[H]
	\begin{center}
		\begin{tabular}{ |p{2.5cm}|p{2.5cm}|p{2.5cm}|p{2.5cm}|  }
 		\hline
 		\multicolumn{4}{|c|}{\textbf{Parameters}} \\
 		\hline
 		\textbf{Parameter} & \textbf{Value} 		&\textbf{Parameter} & \textbf{Value}\\
 		\hline
 		$dt$ & $0.001$ & $T$ & $1$\\
      	$\eta$ & $0$ & $\sigma$ & $0.1$ \\
      	$\bar{\sigma}$ & $0.1$ & $\varsigma$ & $0.1$ \\
      	$q$ & $5$ & dQ & $\pm 1$ \\
      	$\Delta$ & $0.01$ & $\alpha$ & $0.001$\\
            $p$ & $0.2$ & $P_0$ & $50$\\
            $X_{train}$ & $1000$ & $X_{test}$ & $200$\\
 		\hline
		\end{tabular}
	\captionsetup{font=small}
	\caption{Deep RL optimal MM simulation parameters.}
	\label{tab:table1}
	\end{center}
\end{table}

\begin{figure}[H]

\begin{subfigure}{0.5\textwidth}
\includegraphics[width=\linewidth, height=6cm]{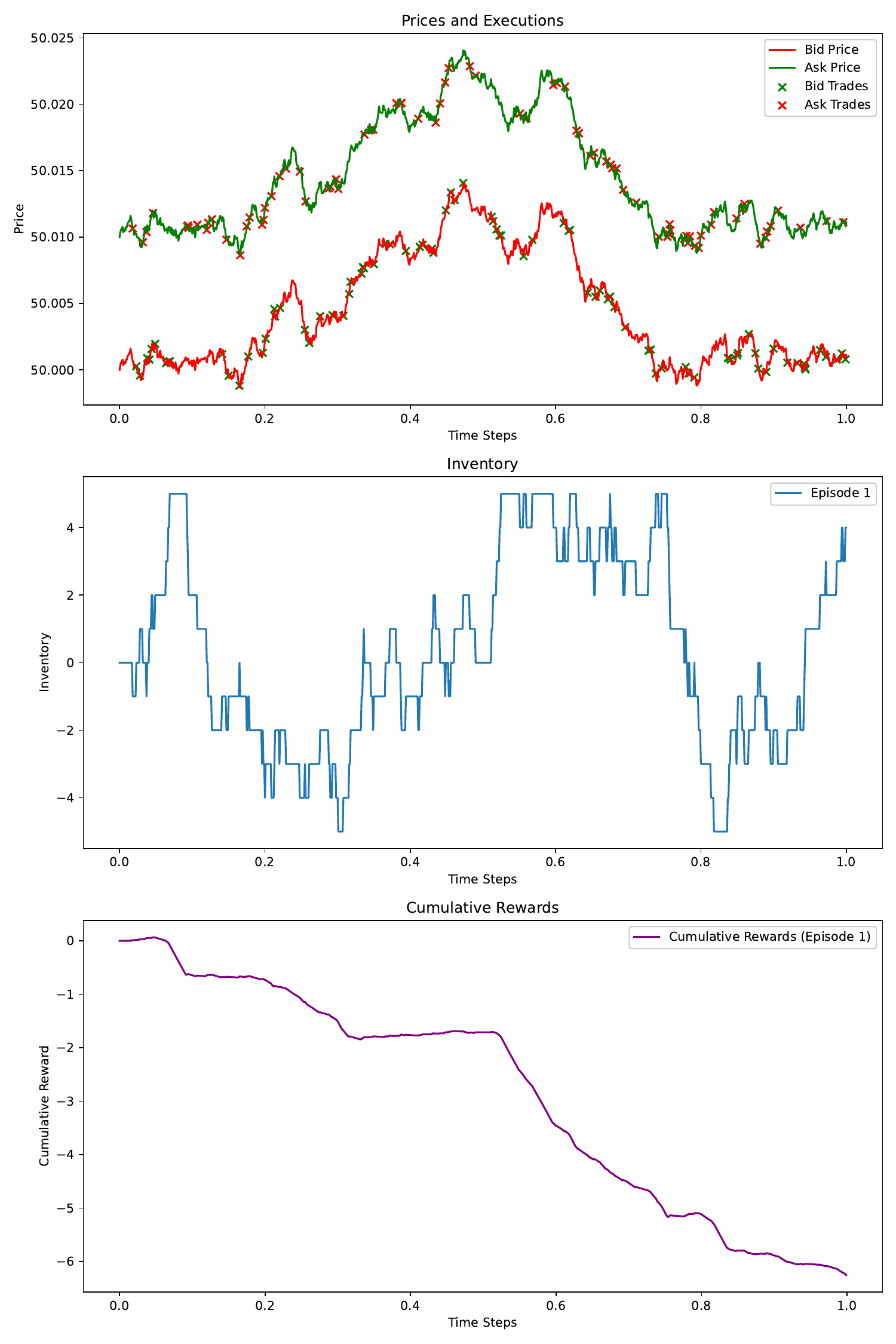} 
\end{subfigure}
\begin{subfigure}{0.5\textwidth}
\includegraphics[width=\linewidth, height=6cm]{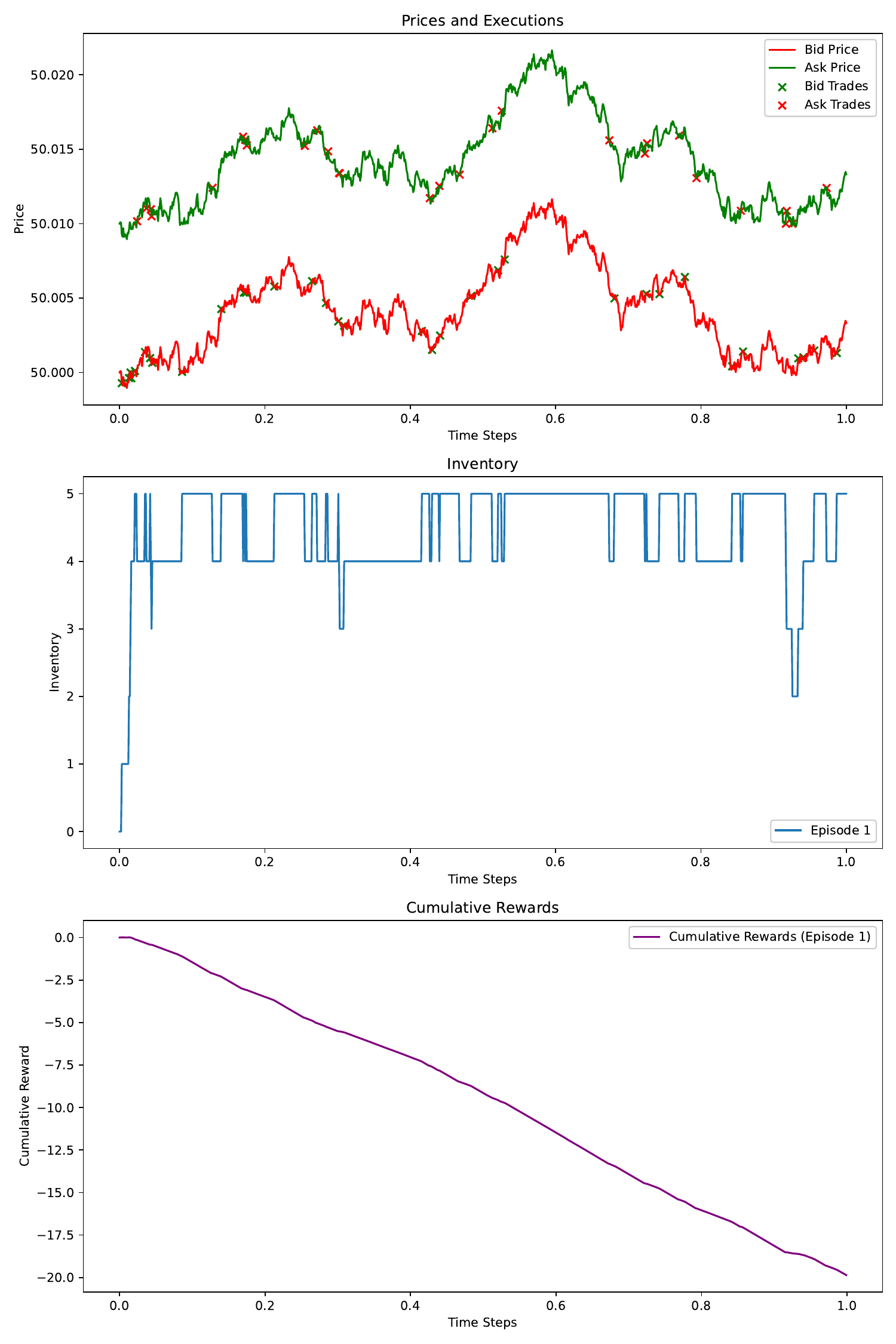}
\end{subfigure}
\captionsetup{font=small}
\caption{Training results from the first training episode, excluding (left) and including (right) adverse fills.}
\label{fig:Training_episode}
\end{figure}

\begin{figure}[H]

\begin{subfigure}{0.5\textwidth}
\includegraphics[width=\linewidth, height=6cm]{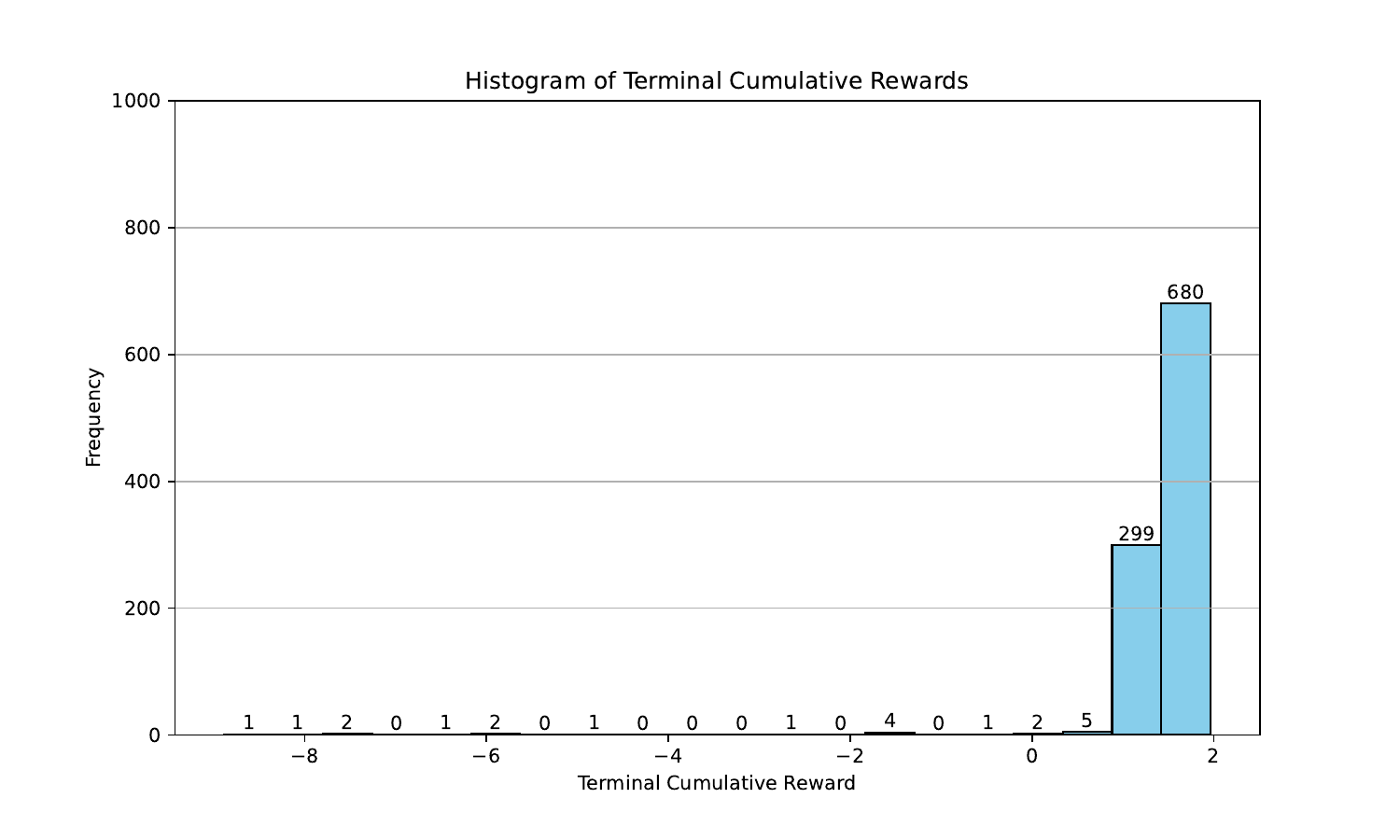} 
\end{subfigure}
\begin{subfigure}{0.5\textwidth}
\includegraphics[width=\linewidth, height=6cm]{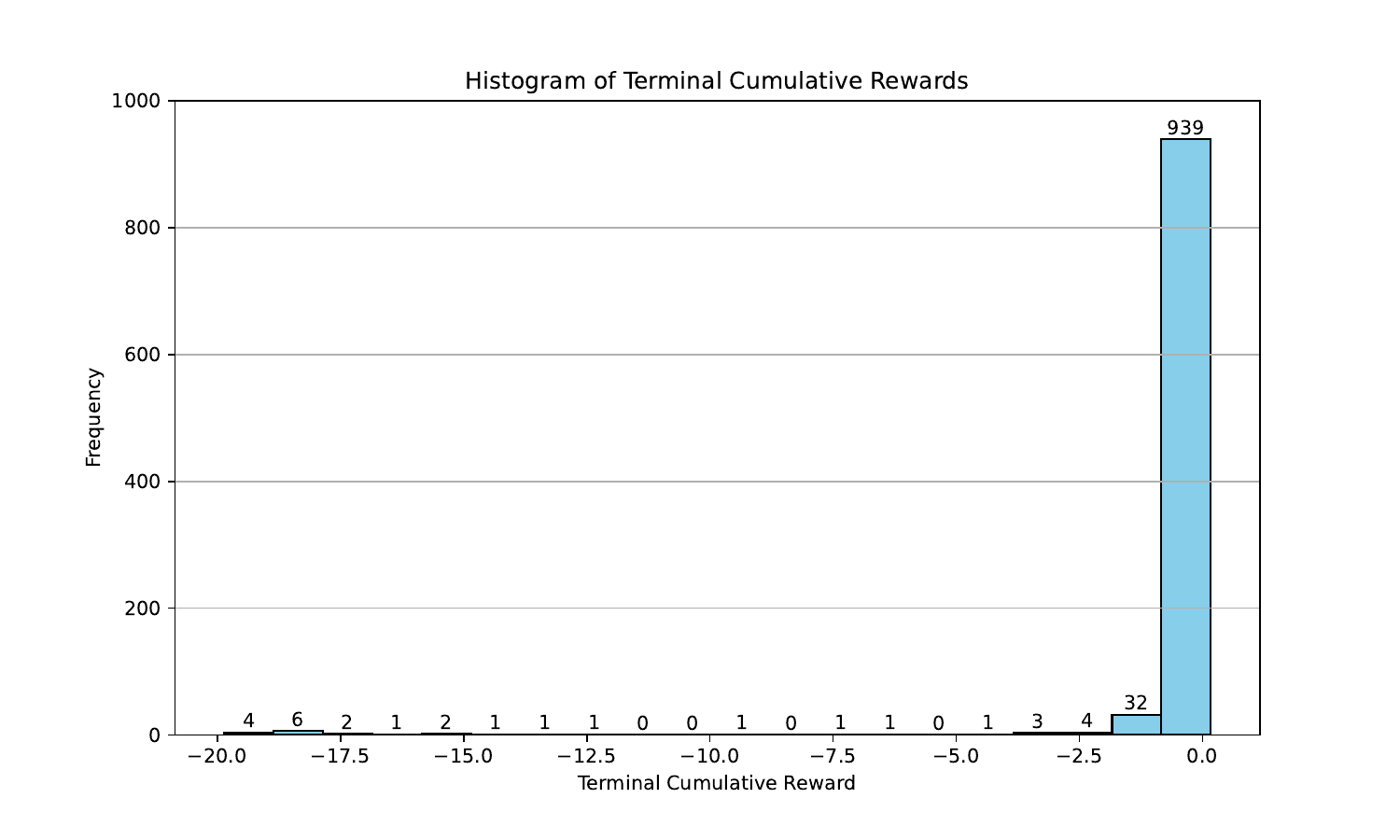}
\end{subfigure}
\captionsetup{font=small}
\caption{Histogram of cumulative rewards over all 1000 training episodes, excluding (left) and including (right) adverse fills.}
\label{fig:Training_rewards}
\end{figure}

To summarize the training results, we provide visuals in figures \ref{fig:Training_episode} and  \ref{fig:Training_rewards}, where the left and right figures portray these results by excluding (left) and including (right) adverse fills. It's quite obvious that including this stylized fact, which is often overlooked in the literature, significantly effects the training results. In the training phase, the agent interacts with the trading environment by taking actions based on the policy. Following every action the MM agent takes, a reward is received after which they transition to the next state. The agents policy is then updated based on its reward function and its state transitions through the SAC algorithm, which we described in section 3. In figure \ref{fig:Training_episode}, we show some results over the first training episode to portray how the optimal MM strategy evolved there. In the top subplots, we show a snapshot of the simulated bid and ask prices, along with the trade executions, then we show the simulated inventories in the second subplot row, and in the last subplot row we show the simulated cumulative rewards. Unsurprisingly, this agent has a negative cumulative reward at the beginning of the training process as it has not been able to figure out the optimal actions to take yet. As we get through more episodes, this slowly improves. To portray this improvement, we show the cumulative results over all 1000 episodes, see the histogram in figure \ref{fig:Training_rewards}, which shows the frequency of every terminal reward over all these episodes. It is quite apparent that the results are mostly positive when adverse fills are excluded, with some heavy left tail results with extreme negative rewards. When adverse fills are included, these results are still quite negative. Thus, although the agent has still improved in the training phase, adverse fills can still negatively hamper the results. This will become more evident when we move to the testing stage next. 

Next, we discuss the results from our test sample, $X_{test}$, which was conducted over 200 episodes. This step is crucial as it helps us determine whether our agent has learned to perform in this type of trading environment and whether a similar performance can be achieved on out-of-sample data. During the testing phase, the agent's policy is fixed, which means it does not update its policy based on any rewards received.  The environment we used under our training and testing sets are the same, which is based on our analysis in section 3, and this ensures unbiased results in machine learning terms. However, to assess the real-world deployment capabilities of any optimal MM problem, it would be prudent to also test the results of the training model in different types of market regimes, in particular regimes that were unseen in the training data, as this could significantly alter the out-of-sample results. Markets regularly change from one regime to another, which can often be measured through metrics such as the implied volatility. One should also test the robustness of any optimal MM strategy like this one, which can be done by changing the parameter values given in table \ref{tab:table1}.

\begin{figure}[H]

\begin{subfigure}{0.5\textwidth}
\includegraphics[width=\linewidth, height=6cm]{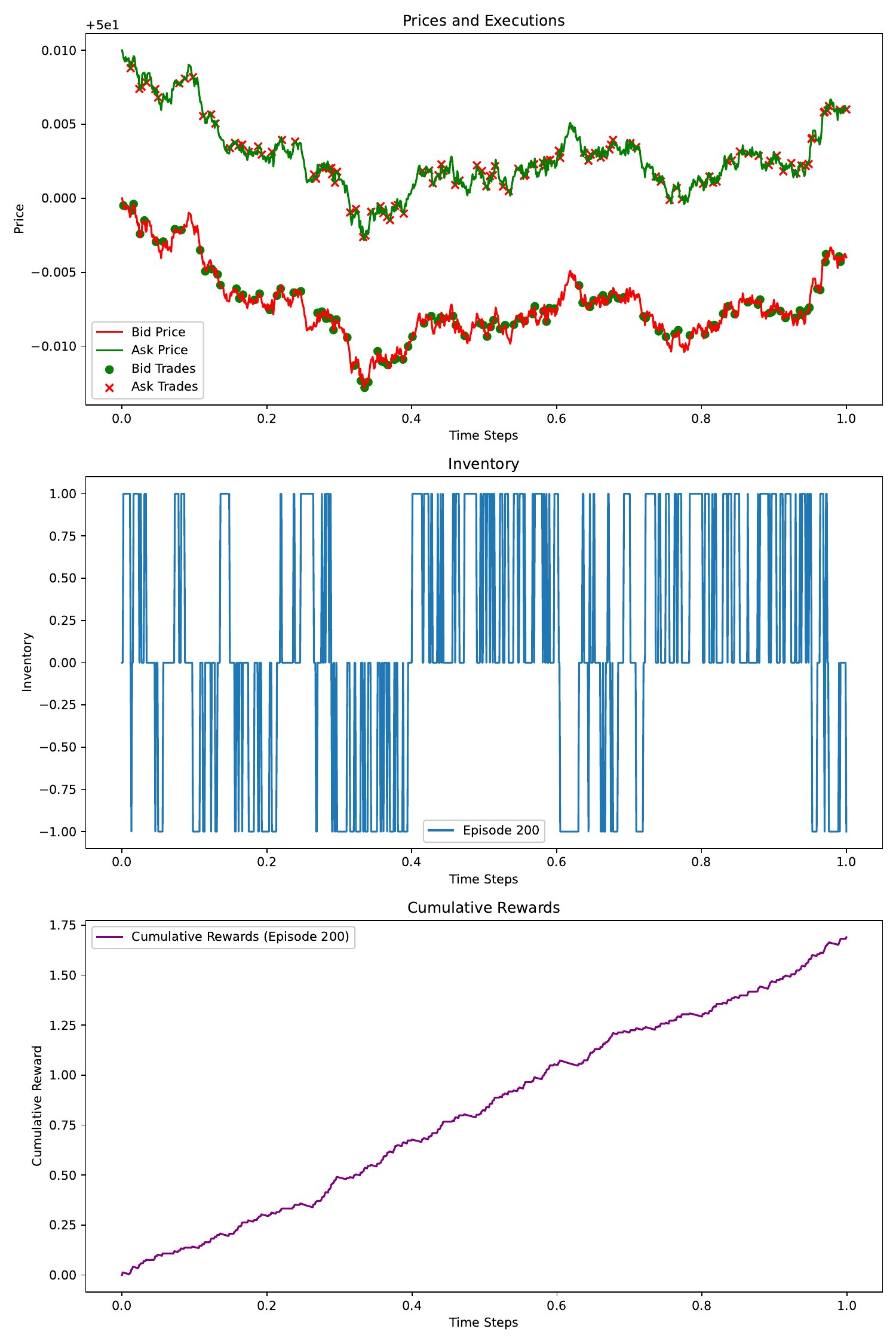} 
\end{subfigure}
\begin{subfigure}{0.5\textwidth}
\includegraphics[width=\linewidth, height=6cm]{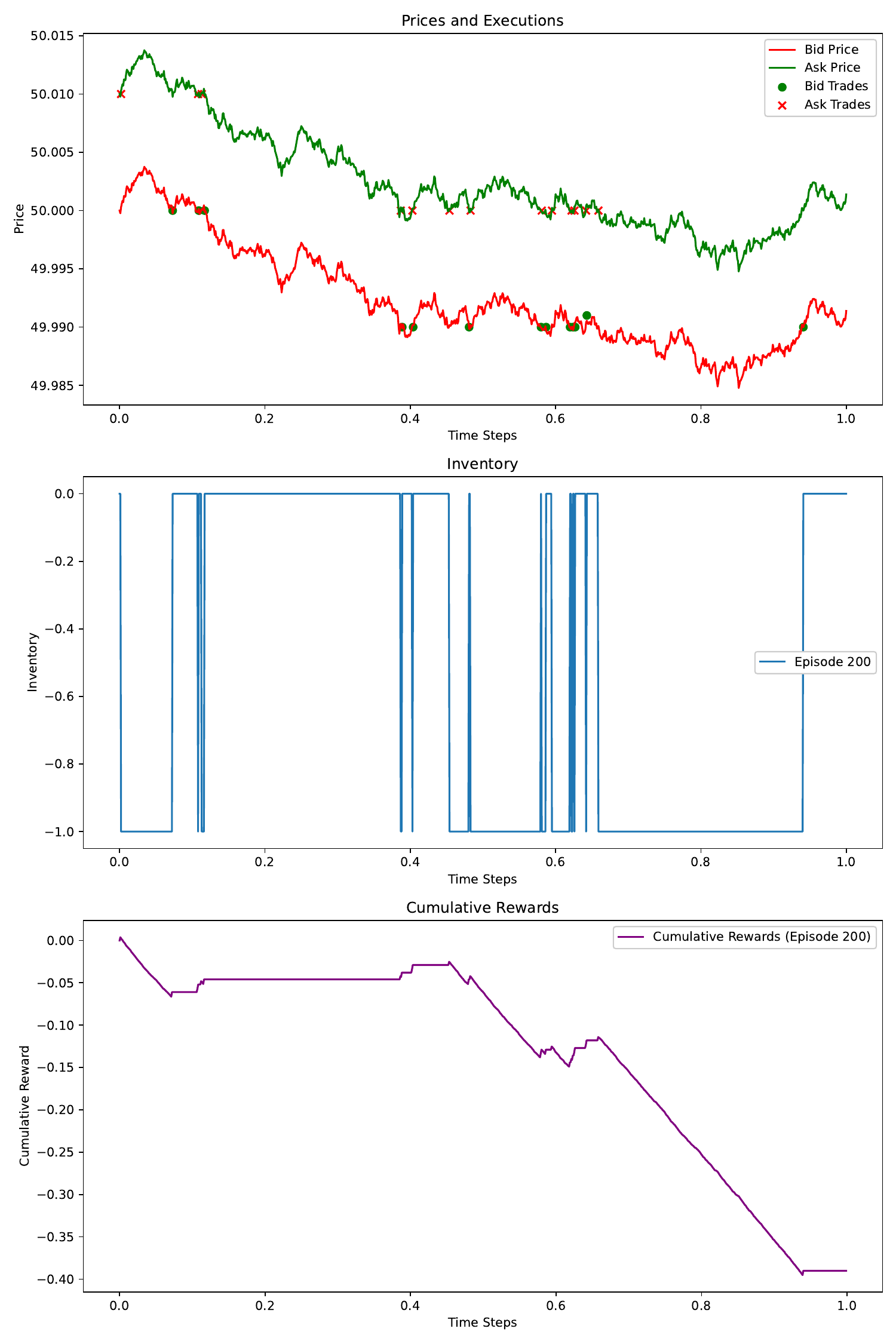}
\end{subfigure}
\captionsetup{font=small}
\caption{Testing results from the last testing episode, excluding (left) and including (right) adverse fills.}
\label{fig:Testing_episode}
\end{figure}

\begin{figure}[H]

\begin{subfigure}{0.5\textwidth}
\includegraphics[width=\linewidth, height=6cm]{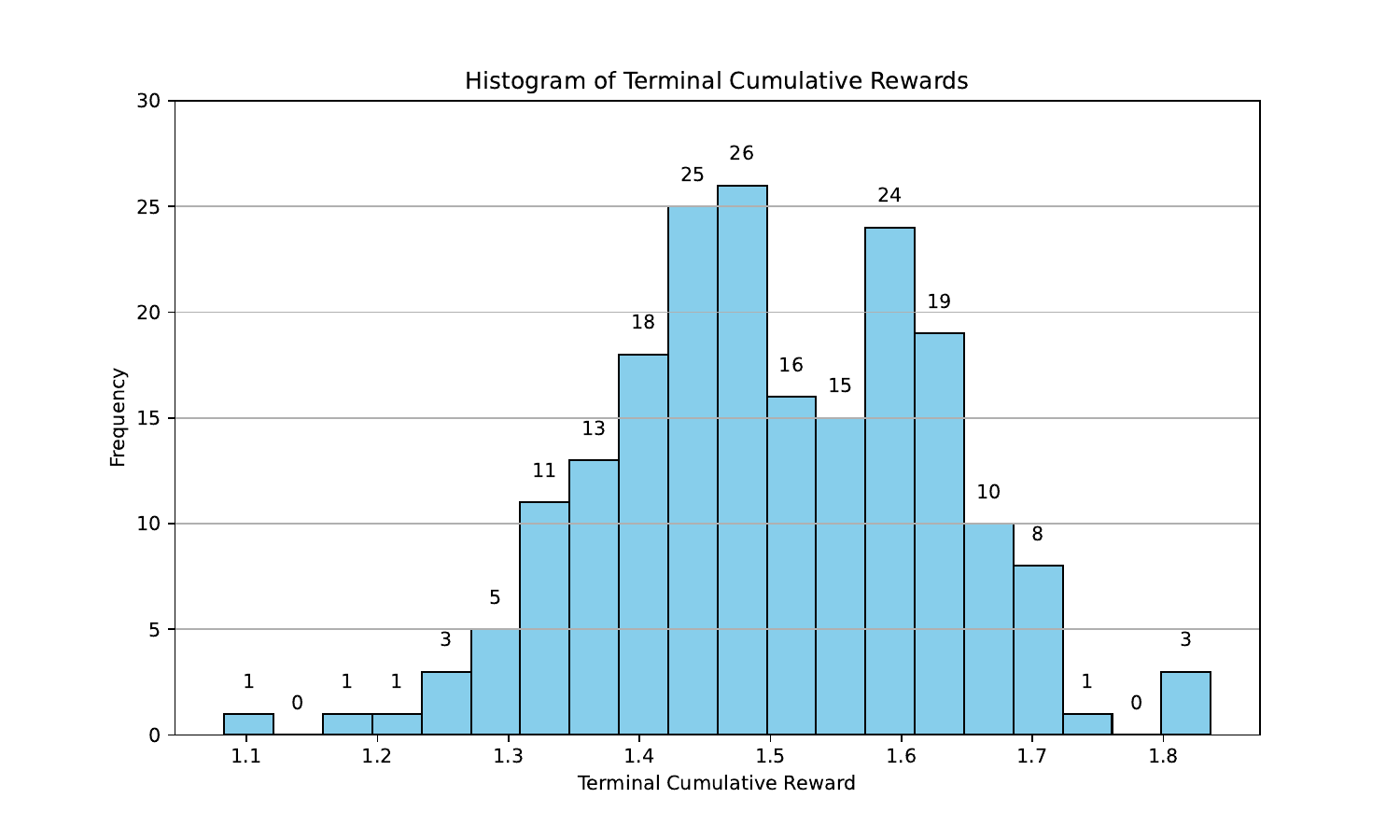} 
\end{subfigure}
\begin{subfigure}{0.5\textwidth}
\includegraphics[width=\linewidth, height=6cm]{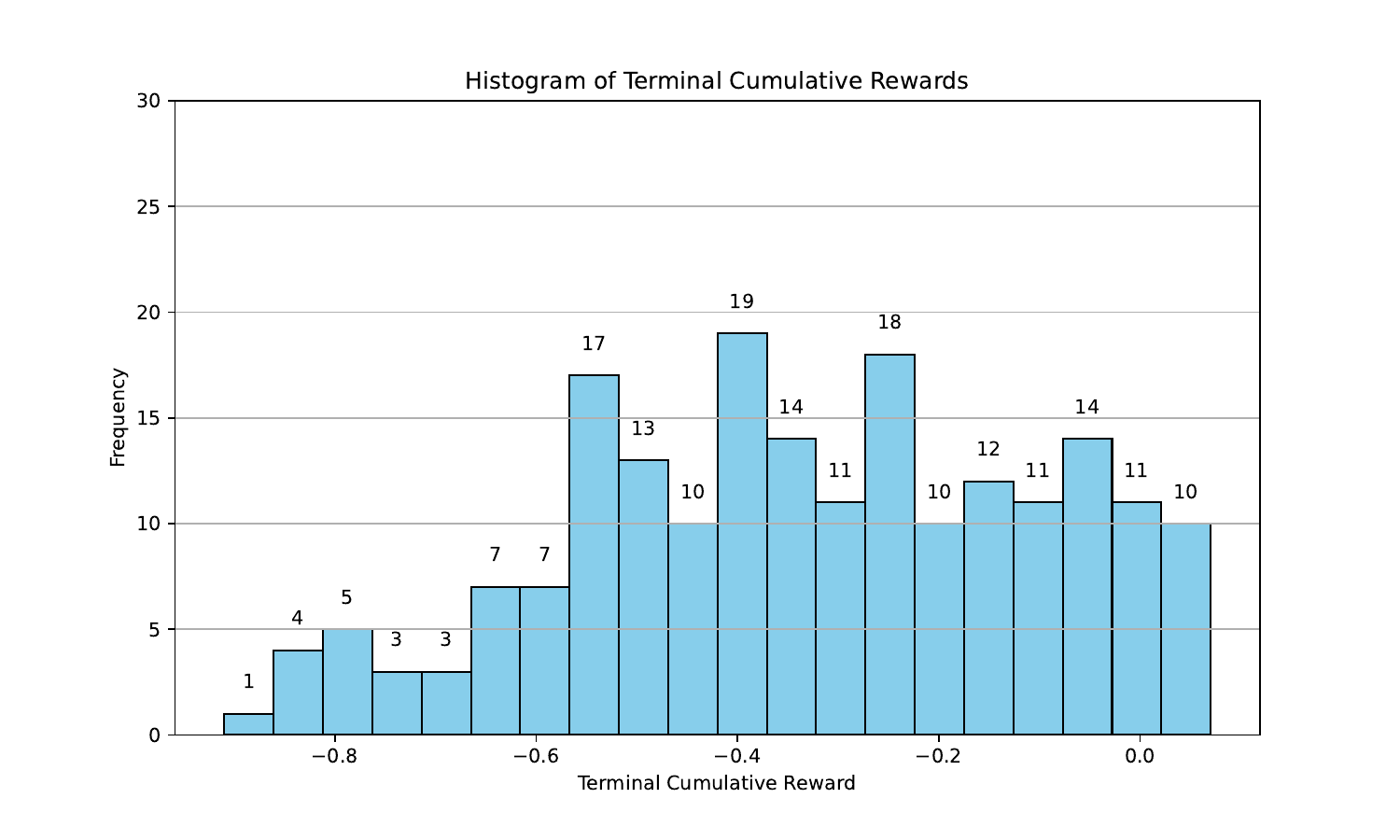}
\end{subfigure}
\captionsetup{font=small}
\caption{Histogram of cumulative rewards over all 200 testing episodes, excluding (left) and including (right) adverse fills.}
\label{fig:Testing_rewards}
\end{figure}

To visualize the testing results, we now provide similar visuals as we just did above for the training results, in figures \ref{fig:Testing_episode} and \ref{fig:Testing_rewards}. It is quiet obvious that the results have significantly improved in comparison to their respective training counterparts (in terms of the exclusion/inclusion of adverse fills), in particular regarding the left tail events present in the training phase, and the histogram shows that the cumulative rewards obtained are much more positive or less negative overall in each episode. In figure \ref{fig:Testing_episode}, we again show a snapshot of the simulated bid and ask prices in the top subplot row, along with the trade executions, then we show the simulated inventory in the second subplot row, and in the last subplot row we show the simulated cumulative reward. One thing that is very apparent is that the deep RL algorithm has learned it is best to not let the inventory deviate from -1 or 1. Thus, keeping a small position is optimal in this scenario. This likely helped to reduce the left tail events as the variables such as adverse fills have less of an effect when they occur if the inventory is low, while this outcome was also likely influenced by the inventory penalty term, $\alpha$. In figure \ref{fig:Testing_rewards}, we can see that the cumulative terminal reward after each episode is much improved, thus under this specific framework the algorithm has learned to consistently generate positive or less negative rewards.

As good as the above results may seem (in terms of learning to generate more positive or less negative cumulative rewards), we would like to stress some of their limitations, as much recent work has shed more light on the limitations of trading strategy back-test results under diffusion pricing models, in particular in a setting where many limit orders are placed, as in this MM strategy. We will now list some of the general limitations, where a great overview was given in \cite{law2019market}. We did try to overcome some of these limitations by making adjustments to our optimal MM strategy, which we will also highlight again. These limitations and adjustments are as follows:
\begin{itemize}
    \item Midprice process: Trading in financial markets generally occurs at the bid and ask, not the midprice. We simulated the midprice at each time step and then assumed a constant 1 tick spread. This might not be too far off reality in some financial markets and some trading regimes (volatility dependent), but often this is not the case and assuming a constant spread could in fact be highly unrealistic in certain markets.  
    \item In these types of frameworks, price and order arrivals are assumed to be independent. In order to tackle this problem, we made sure to include adverse fills in the simulation process as in \cite{lalor2024market}. Excluding adverse fills has the ability to lead to ``large phantom gains", as described in \cite{law2019market}, as MMs as liquidity providers are often on the wrong side of the trade. The empirical results in section 1.1 in \cite{lalor2024market} also shows this, as well through the results we just showed in figures \ref{fig:Training_episode}-\ref{fig:Testing_rewards} above. Thus, including adverse fills has the ability to bring the results a lot closer to reality but still with some limitations, as it doesn't exactly mimic how trading in live markets actually occurs, which is purely discrete based on specific LOB events. An overview of these events, under a reduced-form LOB model, can be seen in table 2 in \cite{law2019market}, which was also applied to the discrete non-diffusive optimal MM deep RL problem in \cite{gavsperov2022deep}. 
    \item The majority of LOBs nowadays use a price-time priority system where orders are filled on a first come first served basis. The models in this analysis and the majority of the HFT literature, however, often just assume that orders are automatically at the front of the queue in the LOB. The models in the literature may use fill probabilities at times but most don't accurately track what the queue position would be in a time-priority market. In order to alleviate this oversight, we utilize the non-adverse fill probability scheme given in \cite{lalor2024market}. While this brings the number of fills a MM would likely receive closer to reality, the timing of how this would occur in reality can still be a bit off. However, it certainly brings the cumulative results of any optimal MM back-test results a lot closer to reality than a situation where this is excluded, which is often seen in the literature. Evidence of this can be found in \cite{lalor2024market}.  
    \item Price ticks: The majority of LOB data assumes only a fixed price grid, normally rounded to two decimal places. In diffusion models, however, this is ignored. Thus, it's illogical to assume a trade transaction could occur at $50.007$ for instance. This can only be rectified by deploying a discrete-time model without diffusion, unless you start rounding midprice values.
\end{itemize}

All in all, the goal of these results was to portray how one could begin to understand the results of deploying a deep RL framework to solve an optimal MM problem. In any trading environment like this, it is crucial to assess the training and testing results, but of course with a grain of salt as in any trading strategy back-test. In order to gain more insight on the subject of accurately assessing strategy back-test results, we recommend reviewing one of the more recent studies on this topic in \cite{arakelian2024discussion}, where they discuss approaches for building a statistically valid back-test. Also, as more and more of the general limitations of diffusion models get resolved, the more applicable these results become in reality. We believe this paper makes strides in the right direction to improving the analysis of an optimal MM framework, particularly under a deep RL framework. 



\section{Conclusions and Future Recommendations}

In this paper, we performed an analysis on an optimal MM problem under a deep RL framework, specifically focusing on semi-Markov and Hawkes Jump-Diffusion pricing models. In our Deep RL framework we applied the SAC algorithm, which is a state-of-the art off-policy method, which often provides more robust solutions to the more complex high-dimensional problems, as in our optimal MM problem. We also implemented numerous steps within the trading environment to better model LOB dynamics. Some of these steps include using non-Markovian pricing dynamics under semi-Markov and Hawkes processes, adverse fills, non-adverse fill probabilities, inventory constraints and inventory penalties. The aim here was to bring a more realistic real-world feel in solving algorithmic and HFT problems and, in turn, results which more accurately mimic what could happen in reality. We do still highlight some of the limitations to these results, as eliminating all of them would make the problem significantly more complicated to setup. Some of these limitations include simulating the midprice dynamics rather than actual bid/ask prices, how trade order fills are being collected and limit order fill probabilities. Of course, these are not the only limitations but some of the main ones we wanted to highlight for the reader who is trying to interpret the applicability of these results. However, we do believe that state-of-the-art frameworks, like in deep RL, have already significantly improved the robustness of trading solutions and the ability to tackle some of these limitations, particularly if you compare them to the standard SOC frameworks like in \cite{bertsimas1998optimal}, \cite{bouchard2011optimal}, \cite{cartea2015algorithmic} and \cite{gueant2017optimal}. 

In terms of recommendations for future research, we would like to highlight that if one could employ a full LOB model into the trading environment, as briefly described in \cite{law2019market}, this would be a big step forward in improving the applicability of these types of models. In \cite{gavsperov2022deep} an attempt is made to create a reduced-form LOB model, where they also cite \cite{law2019market}, which deploys a discrete-time multivariate linear Hawkes process model based on eight of the twelve events in the reduced-form LOB system. Other, and probably more simple model recommendations that we would recommend includes developing models where the spread is not fixed, computing stochastic non-adverse fill probabilities and stochastic volatility coefficients related to all the jump-diffusion coefficients, which may be related to either the semi-Markov or Hawkes models. In terms of the different types of trading problems out there, one could also try to apply this deep RL setting to the trading problems related to Liquidation, Acquisition, Statistical Arbitrage, Volume Imbalance and Pairs Trading as examples, which are some of the heavily studied trading problems under the standard SOC framework in \cite{cartea2015algorithmic}.

\section*{Acknowledgments}
We would like to thank MITACS and NSERC for research funding. 


\bibliographystyle{apalike}
\bibliography{Paper1}

\end{document}